| | |
|---|---|
| Title | **Sustainable improvement of seeds vigor using dry atmospheric plasma priming: evidence through coating wettability, water uptake and plasma reactive chemistry** |
| Authors | T. Dufour, Q. Gutierrez, C. Bailly |
| Affiliations | [1] *LPP, Sorbonne Université, CNRS, Ecole Polytech., Univ. Paris-Sud, Observatoire de Paris, Université Paris-Saclay, PSL Research University, 4 Place Jussieu, 75252 Paris, France*<br>[2] *Sorbonne Université, CNRS, Institut de Biologie Paris-Seine (IBPS), UMR 7622, Biologie du développement, F-75005, Paris, France*<br>E-mail of corresponding author: thierry.dufour@sorbonne-universite.fr |
| Ref. | T. Dufour, Q. Gutierrez, C. Bailly, J. Appl. Phys., Vol. 129, 084902, 2021 |
| DOI | https://doi.org/10.1063/5.0037247 |
| Abstract | Lentil seeds have been packed in a dielectric barrier device and exposed for several minutes to a cold atmospheric plasma generated in helium with/without a reactive gas (nitrogen or oxygen). While no impact is evidenced on germination rates (caping nearly at 100% with/without plasma exposure), seeds vigor is clearly improved with a median germination time decreasing from 1850 min (31h) to 1500 min (26 h), hence representing a time saving of at least 5 hours. We show that the admixture of nitrogen to helium can further increase this time saving up to 8 hours. Contrarily, we demonstrate that the addition of molecular oxygen to the helium discharge does not promote seeds vigor. Whatever the plasma chemistry utilized, these biological effects are accompanied with strong hydrophilization of the seed coating (with a decrease in contact angles from 118° to 25°) as well as increased water absorption (water uptakes measured 8 hours after imbibition are close to 50% for plasma-treated seeds instead of 37% for seeds from the control group). A follow-up of the seeds over a 45-days ageing period shows the sustainability of the plasma-triggered biological effects: whatever the plasma treatment, seeds vigor remains stable and much higher than for seeds unexposed to plasma). For these reasons, the seed-packed dielectric barrier device (SP-DBD) supplied with a He-$N_2$ gas mixture can be considered as a relevant dry atmospheric priming plasma (DAPP) in the same way as those used in routine by seed companies.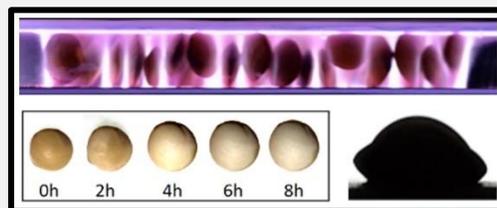 |

# I. State of the Art

## I.1. Improving the germinative properties of seeds using conventional processes

Maintaining seeds quality, procurement and diversity are major challenges routinely addressed through priming processes which consist into imbibing seeds at a controlled temperature and seed moisture content [1]. This partial hydration is achieved to initiate the germination processes without completing it, so that once seeds are about to be sowed by end-users, they can germinate faster, more easily and uniformly. The reason is that the biotic and abiotic stresses associated with the soil and environment that can impact seed germination are mostly alleviated by this technique (e.g. scarcity of water, soil salinity, water logging, deficiency of minerals, extreme temperatures, …).

Several types of priming processes are already worldwide utilized by seed companies, especially hydropriming (seeds soaked in water), osmopriming (seeds soaked in osmotic solutions like polyethylene glycol, urea, $KNO_3$) and halopriming (seeds soaked in salt solutions) [2] [3]. All these techniques are engineered to improve seed germination performances as well as crop yields. As an example, both hydropriming and osmopriming can be successfully applied on melon seeds to alleviate the excess salts in the soil or to mitigate the need for irrigation upon the early stages of plant growth [4]. In the case of lentil seeds, hydropriming reveals to be the simplest technique for enhancing seedling emergence rate. As an alternative to these wet priming approaches, the dry route is also feasible, using for instance solid matrix priming techniques where seeds are incubated in a solid, insoluble and highly water-absorbent carrier (e.g. vermiculite, kieselgur, …) [5]. Drum priming is also a dry route process in which seeds are placed in a rotating drum where water vapor is released to hydrate them appropriately [6].

Although all effective, these techniques face the same limitation which is their processing duration, typically ranging from ten to a hundred hours. Moreover, several technical steps like desiccation may be necessary, hence prolonging the process of several hours and requiring additional manpower. These conventional approaches do not necessarily fit into a logic of sustainable development and their economic costs remain non-negligible. The exploration of disruptive priming technologies must therefore be envisioned to meet the nutritional needs of a growing world population in a context marked by global warming and the need to preserve the environment. From this outlook, dry atmospheric plasma priming (DAPP) appears as an emerging approach in improving germination indicators (vigor, homogeneity, germination, ...), growth parameters (sugar content, nutritive contents, roots dry weight, stems lengths, ...) while taking into account environmental parameters (pathogens loads, $CO_2$ footprint, no toxic chemical residue, …). DAPP could be designed to stimulate the germinative properties of the seeds as well as seedlings growth parameters without the need to go through liquids or long duration treatments.

## I.2. Improving the germinative properties of seeds using dry atmospheric plasma priming

DAPP have already been successfully applied on seeds, inducing a large spectrum of biological effects, from physico-chemical modifications of their coatings to an improvement of their germinative properties and to the subsequent growing properties of seedlings. Besides, decontamination effects of sprouts have also been demonstrated, as detailed hereafter.

### I.2.1. Effects on seeds coating surface

One of the most documented effects is the chemical functionalization of the seeds coating. Plasma has the ability to graft chemical groups likely to trigger various biological effects:
- The easier diffusion of potassium (and other mineral elements of the seed coat) to the inner tissues of the seeds upon water imbibition, as demonstrated by XPS and SEM-EDX on quinoa seeds [7].
- A surface oxidation of the coating. For lentils, beans and wheats seeds, Bormashenko et al. have performed ToF-SIMS analyses which show mass peaks of oxygen and nitrogen ($O^-$, $CNO^-$, $C_2H_3O_2^-$) that are 2.5-3 times more intense on the surface of the plasma-treated seeds than of the non-treated ones [8]. The oxidation of the outermost layers of seed coat and pericarp is also demonstrated on quinoa seeds [7].
- Removal of chemical inhibitors present in the seed envelope [9]
- Covalent or non-covalent attachment of plasma-produced molecular fragments on the seed envelope [9].

DAPP can also be utilized to modify seeds surface topography, either by smoothing their coatings [10] or by texturing/roughening them as in a usual scarification process [9]. In the first case, Ji et al. have demonstrated that seeds of spinach exposed to a micro DBD plasma (atmospheric pressure, 6 kV, 14 mA, 22 kHz, air) result in their layer-by-layer peeling off, even if such mild attack does not explain the improvement of the seeds germinative properties [11]. In the second case, Stolarik et al have showed that cold atmospheric plasmas generate cracks in the external coating of pea seeds [12]. In both cases, high resolution SEM indicate that such physical modifications require a minimum plasma power and a minimum treatment time to be effective. Hence, a low-pressure RF plasma treatment as short as 15 s is insufficient to change the surface topography of lentil, bean and wheat seeds [8].

Tailoring the physical topography and the chemical functionalization of a coating contribute to control its wettability state. The resulting (super) hydrophilic / hydrophobic state can be assessed by depositing a drop of water on the seed and then analyzing its shape through the two contact angles measured on either side of the drop. This wettability state depends only on the first one hundred nanometers thick of the seed coating (sensibility threshold of the technique). Improving this wettability state has already been the subject of several publications. Hence, Bormashenko et al. have shown that bean seeds treated by cold RF plasma exhibited a dramatic hydrophilized surface state with WCA values decreasing from 98° (native) to 53° (plasma-treated) [8]. Complementarily, the same authors have demonstrated that this surface hydrophilization is located on the exotesta of bean seeds (topmost layers of the seed coating) and that the inner tissues (mesotesta, cotyledon) are not affected by the plasma treatment [13]. Usually, this hydrophilization is the result of a chemical functionalization of oxygenated/nitrogen functions (e.g. –OH, –O, –OOH, …) even if, as underlined by Randeniya, inert gas plasma alone can also improve the hydrophilic state of seeds coatings [14]. Such unexpected observation may result from: (i) a residual background of air in the plasma device if no vacuum process is carried out beforehand, (ii) the action of plasma on the moisture initially contained in the seeds, thus leading to the production of key radicals involved in the promotion of seeds germinative properties [14]. So far, no obvious correlation has been found between the strong plasma-triggered hydrophilization of the most superficial layers of the coating with the low rise in water uptake [8] [7].

In addition to the surface wettability issues, the question of coating permeability appears quite relevant even if it remains poorly addressed in the literature. Under saline and osmotic stress conditions, Bafoil et al. have applied the same plasma treatment (DBD in ambient air) on *Arabidopsis thaliana* seeds. Whatever the genotype of these seeds (Col-0, gl2 and gpat5), the authors manage to reduce coating permeability, hence making the seeds less sensitive to the inhibitory effects of their environment. This lower permeability is explained as the result of two types of plasma-triggered surface modifications: (i) the morphological changes like etching and smoothing and (ii) the exudation of lipid compounds which can subsequently modify the surface chemical composition [15]. The works of Volin et al. deserve also a special interest: radish, corn, soybean and pea seeds are exposed to an RF plasma in a low-pressure rotating reactor. Various organic/inorganic macromolecules are injected as vapor or gas [16]. Fluorocarbon plasmas like $CF_4$ or ODFD (octadecafluorodecalin) leads to the deposal of hydrophobic films (5 µm in thickness) on the seeds, then resulting in a delay of the vigor [16]. The hydrophobic property of the deposited polymer film as well as the induced osmotic stress are both considered as responsible for such reduction in germination speed.

### I.2.2. Effects on seeds germinative parameters

In the case of non-dormant seeds, several works have already shown that DAPP can increase seeds vigor without raising their germination rate. Hence, with an inductive air plasma (2 min, pressure lower than 0.1 Pa), it is possible to decrease the median germination time of bean seeds from 44h (control) to 40h (plasma) while maintaining a germination rate close to 95% in the two groups [13].

Regardless the issues of seeds dormancy release, many articles have reported the benefits of DAPP in surging the germination potential. One of the most spectacular result concerns *Paulownia tomentosa* seeds: while it remains as low as 5-30% for untreated seeds, it reaches a threshold as high as 75% after a low-pressure air plasma (5 min, 200 mTorr) [9]. In the case of *C. tinctorium L. semen* seeds, germination rate increases from 12% (control) to 18% (low-pressure RF argon plasma , 16 Pa, 20 W) [10]. Besides, RF low-pressure plasma (13.56 MHz, 80W, 150 Pa, He) applied on tomato seeds leads to a higher germination rate (80% for control vs 91% for plasma) [17]. However, plasma can also induce reverse effects: spinach seeds exposed to a high voltage nanosecond pulsed plasma (atmospheric pressure, 6 kV, 700 A, 1-10 shots) are severely surface-degenerated with increasing number of shots, driving to a decay of both the germination rate and subsequent seedling growth [11]. The mechanisms ruling the effects of plasma on germination rate enhancement are partially addressed by Stolarik et al. Their cold plasma treatment (Coplanar-DBD, 1-10 min, 1 atm) slightly increases the water uptake of pea seeds, promoting a faster hydratation and activation of germination enzymes and phytohormones (auxins and cytokinins as well as their catabolites and conjugates) [12]. Their DAPP strongly increases the levels of dihydrozeatin riboside (+98.6% compared to the control) and dihydrozeatin riboside-O-glucoside (+179% compared to the control) [12]. If the germination rates increase significantly (from 77.5% to 82.5%), the seedlings growth parameters remain almost unchanged, e.g. root length rising from 12.03 cm (control) to 12.33 cm (plasma) [12].

Finally, some research works indicate that plasma can both improve germination potential and vigor. After exposure to a low-pressure discharge, seeds of Lamb's Quarters evidence a rise in the germination rate from 15% (control) to 55% (plasma), accompanied with a significant acceleration of the germination time [18]. Oats seeds exposed to a low-pressure plasma process (0.1 Torr) germinate six days earlier than the untreated seeds while germination rate is 27% higher when the plasma operates in a DC regime rather than a pulsed regime [19]. However, no plasma-driven post-germination effects are evidenced, e.g. at day 11, the length of seedlings from control and plasma groups remain almost the same (14.5 cm) [19].

### I.2.3. Effects on seedlings growth parameters

Some experimental work show that plasma-treated seeds may have negligible effects on subsequent parameters of shoot growth. Hence, maize seeds exposed to a dielectric barrier discharge in ambient air has no effect on root anatomy or morphology, and does not drive to change in enzymatic activities (G-POX, SOD and DHO) except catalase activity which is measured over a 1 week follow-up [20].

However, most research works attest that seeds exposed to plasma undergo anatomical and morphological modifications of sprouts. Such modifications are correlated with changes in various enzymatic and hormonal activities:
- Radish seeds exposed during 20 minutes to a DBD (15 kV, 50 Hz) exhibit biological effects at early stages of seedlings growth while roots lengths and roots weight increase by 10% and 30% respectively [21]. This observation is confirmed by the works of Matra et al. where sprouts of radish exposed to a cold atmospheric plasma of argon have seen an increase of 9-12% in dry weight [22].
- In the case of tomato seeds treated with a low-pressure magnetized plasma, the POD activity in the seedling's hypocotyls is markedly increased compared with control (860 u.mg$^{-1}$.pro.min$^{-1}$ vs 410 u.mg$^{-1}$.pro.min$^{-1}$ respectively), as well as the hydrogenase activity of the roots assessed through the amount of TTC reduction (2.3 mg.root$^{-1}$.h$^{-1}$ for control vs 5.9 mg.root$^{-1}$.h$^{-1}$ for plasma) [23]. Another low-pressure DAPP (13.56 MHz, 80W, 150 Pa, He) performed on tomato seeds result in seedlings with activities of POD (421.3 U gFW$^{-1}$), PPO (508.8 U gFW$^{-1}$) and PAL (707.3 U gFW$^{-1}$) greater than the ones measured in the seedlings' controls (103.0 U gFW$^{-1}$, 166.0 U gFW$^{-1}$ and 309.4 U gFW$^{-1}$, respectively) [17]. As a result, growth parameters like plant height increase from 49.0 cm (control) to 54.3 cm (plasma) while more calcium and boron are absorbed. Tomato seeds have also been treated using a DBD reactor operating in ambient air at atmospheric pressure (15 slm, 17 kV, 50 Hz). If such treatment does not improve that much the germination rate, it considerably influences growth parameters (e.g., mean root length at day 6 is 45 mm long for control vs 66 mm long for plasma) [24]. Besides, an increase in weight of 20-40% is observed for the plants grown from treated seeds as compared to the control ones [24]. The best results are obtained for relatively short plasma exposure, while – surprisingly – longer DAPP damage the seeds.
- Seeds of wheat treated by a low-pressure plasma process (He, 80W, 150 Pa, 3 GHz) see an increase in their germinative parameters (e.g. germination rate 6.7% higher compared with control) as well as in growth parameters: plant height (+21.8%), root length (+11.0%), stem diameter (+9.0%), leaf area (+13.0% and leaf thickness (2+5.5%), chlorophyll content (+9.8%) and nitrogen (+10.0%) [25]. Complementarily with this work, Sera et al. show that after treating wheat seeds with low pressure plasma (500W, air, 0-2400s, 140 Pa), the phenolic compounds of the seedlings could significantly increase, hence suggesting changes in metabolism processes as well as the penetration of active species from the plasma through the porous seed coat to react with seed cells [26]. Wheat seeds are also treated at atmospheric pressure with a surface DBD operating at room temperature. Here again, wheat seeds exhibit enhanced growth parameters: the mean root weight raises from 0.78 g to 1.06 g while curve-fitted gaussian distributions of sprout lengths are two times narrower compared to control [27]. As a result, the root-to-shoot ratio rises from 0.9 to 1.2.
- In the case of oilseed rape treated with low-pressure plasma process (13.56 MHz, 100W, 150 Pa, 15 min, 25°C), the germination rate increases by 6.25%, accompanied with a raise in superoxide dismutase and catalase activities by 17.71% and 16.52% respectively, as well as increase in protein contents and in soluble sugar which is known to be the main form of carbohydrate and the photosynthetic product of plants [28]. Therefore, by upregulating antioxidant enzyme activities, improving osmotic adjustment and decreasing oxidative lipid degradation, DAPP could be used to make seedlings more resistant to various stress conditions like drought [28].

- After plasma treatment (low-pressure process, 3 kV, 10 min), the alpha-amylase activity measured in germinating brown rice increases from 1.5 to 4.5 mg glucose/g dry wt. while increased levels of gamma-aminobutyric acid (GABA) are obtained (from 19 mg/100 g for control to 28 mg/100 g for plasma) [29]. The authors claim that such higher enzyme activities are responsible for triggering the rapid germination and earlier vigor of the seedlings [29]. In the case of soybean seeds exposed to cold plasma (15s, 13.56 MHz, 80 W, 150 Pa) germination rate and vigor indexes increase by 14.7% and 63.3%, respectively, while soluble sugar, protein contents, shoot weight and root weight are 16.5%, 25.1%, 21.9% and 27.5% higher than those of the control respectively [30].

### I.2.4. Effects on seeds and sprouts decontamination

Whether during their cultivation, preparation, transport or storage, seeds and sprouts are likely to be contaminated by various toxic agents (e.g. heavy metals, endocrine disruptors, radionuclides, pesticide residues, etc.) as well as micro-organisms (ex: *salmonella, E. coli, Aspergillus spp.,* ). In particular, the retail sale of sprouts poses food safety concerns, first because they are cultivated in a very bacteria-friendly environment and second because they are eaten raw. For plasma agriculture, the challenge is therefore to find the right balance between inactivating the seeds pathogens to high decontamination levels while preserving/improving the seeds germinative properties.

In this context, Jiang et *al.* have demonstrated that seeds of tomatoes exposed to cold plasma (13.56 MHz, 80W, 150 Pa, He) give seedlings with stronger resistance against *R. solanacearum*: 20 days after bacterial inoculation, disease severity is 100% for control vs 71% for plasma [17]. According to Brasoveanu et al., seeds decontamination levels depend on the types of fungi: in the case of a low-pressure DAPP (air, 15 Pa, 100-200W, 20 min), the total number of fungi decrease only from 220 a.u. to 170 a.u. for barley seeds and from 220 a.u. to 140 a.u. for corn seeds [31]. Slightly undesirable effects are also obtained on the germination rates since they decrease from 95% (control) to 88% (plasma) for barley seeds and from 98% (control) to 95% (plasma) for corn seeds [31]. Fortunately, such adverse effects do not appear in the works of Selcuk et al. where, in the case of various seeds (wheat, barley, oat, lentil, rye, corn, chickpea), the low-pressure plasma (air/$SF_6$, 20 minutes) has no impact on the initial germination rates while fungi loads of *Penicillum spp.* and *Aspergillus spp.* are reduced between 1 and 3 decades [32].

Increasing the working pressure to 1 atmosphere is not a technological lock for plasma decontamination: seeds of *Cicer Arietinum* exposed during 1 min to the FlatPlaSter plasma device (ambient air, atmospheric pressure) drive to a 1-log reduction of the natural microbiota (from 4.5 to 3.6 log CFU.$mL^{-1}$.$cm^{-2}$). In addition, a higher germination rate and an accelerated germination time (2.7 days instead of 3.5 days) are observed [33]. Results even more encouraging are obtained with cress seeds treated with an atmospheric pulsed DBD device (10 min, 8 kV, 10kHz, 500 ns pulse) with a reduction of 3.4 log of *E. coli* [34].

### I.2.5. Effects induced by DAPP as a function of ageing time

Ageing effects resulting from seeds exposed to DAPP are poorly investigated in the literature. In the case of *Arabidopsis thaliana* seeds, Bafoil et al demonstrated that the plasma generated by an FE-DBD is responsible for a testa failure rate of at least 5% compared to untreated seeds (control) , and that this increase is preserved from 1 to 9 days after plasma exposure. At the same time, these same plasma-treated seeds show an initially 15% higher endosperm rupture rate compared to the control. However, after 9 days, this rate drops to the same level as that of the control group [35]. The issue of sustainability is also addressed in the works of Meiqiang et *al.* where ageing effects are discussed considering tomato seeds treated with a magnetized plasma device. It turns out that over a harvest period of 26 days, plants originating from the plasma-treated seeds exhibit crop yields higher than 20% compared with untreated seeds. However, ageing effects are evidenced for second harvest achieved between days 26 and 78 since crop yields are reduced to roughly 20% in comparison with control [23].

## I.3. Investigating how dry plasma can improve the germinative parameters of seeds in a sustainable way

In the present work, we investigate the effects of DAPP on seeds by changing the chemical composition of plasma through $N_2$ and $O_2$ gases admixed with helium. We propose to evidence the effects of such dry atmospheric plasma primings on vigor, homogeneity and germination rate of lentil seeds. In this outlook, a correlation is achieved between (i) the relative densities of actives species within the plasma phase (through optical emission spectroscopy and mass spectrometry measurements) and (ii) seeds parameters related to hydric properties (water uptakes and water contact angles of coatings). All these experiments are achieved at different ageing times over a 45-days period.

Lentil is chosen as the reference agronomical model of this article owing to its agro-ecological and nutritional assets as well as its yearly production rate as high as 3.6 million tons worldwide [36]. If lentil appears as one of the main food legumes, its crops yields is jeopardized by various (a)biotic stresses causally related to global warming, especially drought, heat, insect pests and diseases [36]. The present work envisions dry atmospheric plasma technology as a relevant tool to promote the germinative parameters of seeds sowed under (a)biotic stress conditions, whether mild or harsh.

# II. Experimental setup, materials and methods

## II.1. Germination follow-up model

The proportion of germinating seeds after their imbibition in water (t=0), follows an exposure-response model. A method for mathematically describing such phenomenon relies on the utilization of the four-parameter Hill function, also expressed as a dose-response curve with variable Hill slope [37] as given by equation (1):

$$G_\%(t) = \frac{G_\%^{max} - G_\%^{min}}{1 + 10^{p.(\tau_{50}-t)}} \quad (1)$$

where t is time, $G_\%(t)$ the proportion of germinated seeds at a given instant, $G_\%^{min}$ the initial proportion of germinated seeds (i.e. 0 seeds), $G_\%^{max}$ the maximum germination rate reached at the end of the experiment, p the slope, $\tau_{50}$ the median germination time corresponding to the reciprocal function of $G_\%(t)$ at $\frac{G_\%^{max}-G_\%^{min}}{2}$. As a physiological process, germination can be assessed following three indicators: $\tau_{50}$, $G_\%^{max}$ and S which stand for vigor, germination potential and homogeneity respectively. As reported in Table 1, these indicators can also be expressed considering seeds from control group as a reference.

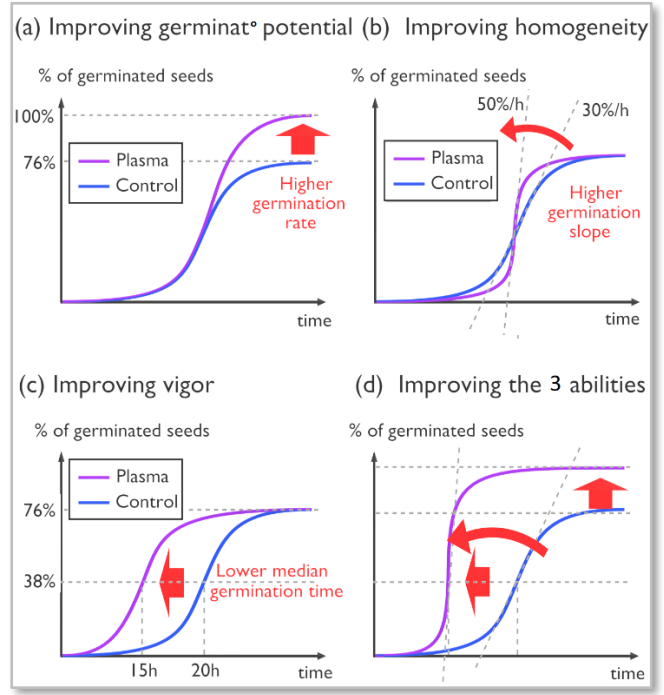

*Figure 1. Sketch diagrams of germination curves considering improvements of (a) germination potential, (b) homogeneity, (c) vigor, (d) the three aforementioned abilities.*

| Ability | Germination indicators | | Units |
|---|---|---|---|
| | Absolute | Relative | |
| Vigor | Median germination time $\tau_{50} = G_\%^{-1}\left(\frac{G_\%^{max} - G_\%^{min}}{2}\right)$ | $\Delta\tau = (\tau_{50})_{ctrl} - (\tau_{50})_{pl}$ | min |
| Germination potential | Germination rate $G_\%^{max} = G_\%(t = t_{max})$ | $\Delta G = (G_\%^{max})_{ctrl} - (G_\%^{max})_{pl}$ | % |
| Homogeneity | Germination slope $S = p.Ln(10).G_\%^{max}$ | $\Delta S = (S)_{ctrl} - (S)_{pl}$ | %.min$^{-1}$ |

*Table 1. Indicators for monitoring seed germination abilities.*

Whether relying on plasma or any other physico-chemical approach, a priming process is intended to meet three main stakes: increasing germination rate, increasing germination homogeneity, increasing the vigor and – ideally – improving these three indicators altogether, as sketched in Figure 1.

## II.2. Plasma source

Seeds of lentils (*Lens culinaris*, from *Radis & Capucine* company) have been packed in a dielectric barrier device constituted by a rectangular quartz tube (30 cm length, 6 mm × 15 mm inner section) platted by two aluminum electrodes. As sketched in Figure 2, the resulting interelectrode gap (6 × 15 × 50 mm³) can contain approximately 80 seeds of lentils (the unitary volume of one seed being 23.84 mm³). The resulting seed-packed dielectric barrier device (SP-DBD) is powered with a high voltage generator composed of a function generator (ELC Annecy France, GF467AF) and a power amplifier (Crest Audio, 5500W,CC5500). The SP-DBD is supplied with 6 kV in amplitude at 600 Hz (sine form) and filled with helium gas (2 slm) mixed with/without molecular nitrogen or oxygen (0-150 sccm) at atmospheric pressure. Before switching on the plasma, a purge procedure is carried out during which helium gas is injected for 2 minutes to drastically reduce residual air background.

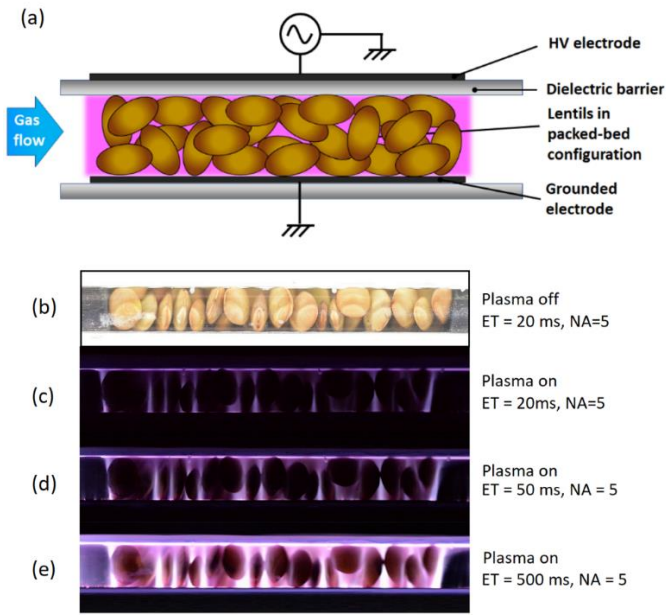

*Figure 2. (a) Sketch of the seed-packed dielectric barrier device (SP-DBD) operating at atmospheric pressure (b) Picture of the SP-DBD when plasma is off, (c-e) pictures of the SP-DBD when plasma is ignited with helium at 2 slm, 6 kV, 600 Hz, considering various exposure times (ET) but same numerical aperture (NA)*

## II.3. Diagnostics

Electrical characterizations of the SP-DBD have been achieved with high voltage probes (Tektronix P6015A 1000:1, Teledyne LeCroy PPE 20 kV 1000:1, Teledyne LeCroy PP020 10:1) and an analog oscilloscope (Wavesurfer 3054, Teledyne LeCroy).

The gaseous phase has been analyzed using a quadrupole-based mass spectrometer (Model HPR-20 from Hiden Analytical Ltd). Plasma chemical species have been collected by a quartz capillary, 1 m in length, flexible, chemically inert and heated at 200 °C to prevent chemisorption. Then, a three-stage differentially pumped inlet system separated by aligned skimmer cones and turbo molecular pump, enables a pressure gradient from $10^5$ bar to $10^{-7}$ bar at the entrance of the ionization chamber. There, ionization energy is set at 70 eV. The residual gas analyzer (RGA) detector is used for scanning masses from 1 to 50 amu.

The radiative species from plasma have been detected using an optical emission spectrometer from Andor (SR-750-B1-R model) equipped with an ICCD camera (Istar model). The spectrometer operates in the Czerny Turner configuration with a focal length of 750 mm while diffraction is achieved with a 1200 grooves.mm$^{-1}$ grating in the visible range. The following parameters have been selected for all experiments: intensification factor = 2000, exposure time = 100 ms, number of accumulations = 70.

Water contact angles have been measured on seeds teguments to assess their wettability state. The Sessile drop technique is followed using a drop shape analyzer engineered and assembled at laboratory. All contact angle measurements are deposited with drops of 10 µL in volume (milli-Q water). In this work, each value of contact angle corresponds to an average of 10 drops deposited on 10 different seeds.

The amount of water that a seed can absorb is evaluated by measuring the mass of that seed before and after its imbibition. The resulting water uptake can be defined by equation (2) from Bormashenko et al [08]:

$$\xi(t) = 100 \times \frac{m(t) - m_0}{m_0} \quad (2)$$

where $m_0$ corresponds to the initial mass of the seed lot in the dry state (here 1 lot = 30 seeds) and $m(t)$ corresponds to the mass of this same lot at a post-imbibition time t. The weighs have been measured using an analytical balance (SARTORIUS company, model Entris 124i-1S) with readability of 0.1 mg and reproducibility ±0.1 mg.

## II.4. Methods for germination follow-up

Germination assays have been carried out following the recommendation of FAO for germination testing [38]. The effects of plasma have been systematically evaluated on 5 independent samples of 80 seeds, i.e. 400 seeds in total. The following procedure has been followed:

(i) After plasma treatment, samples are stored for a $t_{ageing}$ period corresponding to 0, 15, 30 or 45 days. Seeds are stored in the dark at a temperature of 20 °C (± 0.5 °C) and a relative humidity (RH) of 40 % (± 2%).
(ii) Commercial squares of cotton wool (8cm × 8cm × 3mm) are utilized as test substrates and placed in Petri dishes. Each substrate is soaked with 10 mL of tap water. Its chemical parameters, reported in Table 2, are calculated from the sanitary control agency of the water distribution network in Paris [39].
(iii) Distribution of each sample (80 seeds) in batches of 20 seeds per Petri dish. To prevent any potential spread of fungal moulds, seeds are spaced the ones from the others uniformly by a distance corresponding to 2.5 times their diameters.
(iv) Storage of Petri dishes in the dark at 20 °C and RH = 40%, throughout the germination follow-up.

| Parameter | Average |
|---|---|
| Turbidity | 0.4 NFU |
| Free chlorine | 0.1 mg($Cl_2$)/L |
| Conductivity | 590.8 µS/cm |
| pH | 7.6 |
| Iron | 9.4 µg/L |
| Nitrates | 47.8 mg/L |
| Ammonium | 0.0 mg/L |
| Escherichia Coli | 0 n/(100mL) |
| Hardness | 29.2 °f |
| Calcium | 108.1 mg/L |
| Bicarbonates | 295.0 mg/L |

*Table 2. Composition of the tap water utilized for seeds imbibition.*

# III. Results & Discussion

## III.1. Characterization of plasma in a seed-packed dielectric barrier device

Since no vacuum purge procedure is achieved in the dielectric barrier device before plasma-processing seeds, a residual background of ambient air remains present even if the device is only supplied with helium. As shown in Figures 3a and 3b, when no reactive gas is admixed with helium, mass spectrometry indicates the presence of water molecules roughly 2 decades lower than the He intensity as well as molecular nitrogen and oxygen 3 decades lower. In Figure 3a, the admixture of $N_2$ from 0 to 150 sccm leads unsurprisingly to a higher $N_2$ intensity as well as a low increase of the nitric oxide signal while He, $H_2O$ and $O_2$ intensities remain constants. In Figure 3b, the admixture of molecular oxygen to the 2slm flow rate of helium has no influence on the intensities of helium, $H_2O$ and $N_2$ but drives to a clear increase of the $O_3$ signal correlated with the increasing $O_2$ intensity. It is noteworthy to remind that short lifetime species like OH radicals are subject to recombination in the MS capillary and cannot be detected. Optical emission spectroscopy has therefore been achieved to supplement these results.

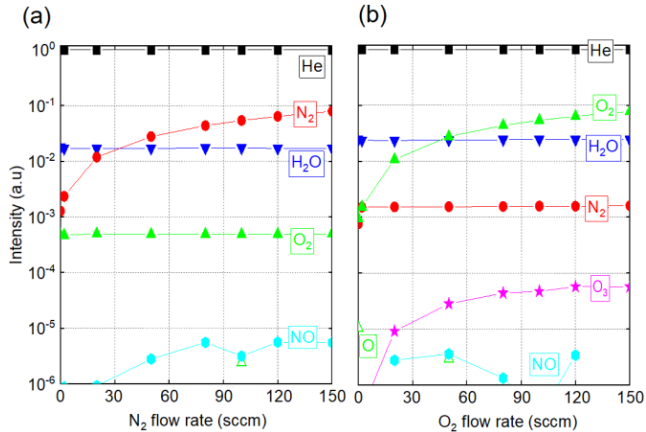

*Figure 3. Intensity of plasma species detected by mass spectrometry as a function of (a) $N_2$ flow rate or (b) $O_2$ flow rate. In all cases, the reactive gas ($N_2$ or $O_2$) is mixed with 2 slm of helium; all measurements are achieved in SP-DBD configuration.*

The emission of the electronically excited states of plasma species is proportional to their population density characterized by (i) the electron energy distribution function and (ii) the density of the plasma species involved in the optical emission. The Figure 4a introduces the energetic levels and transitions of the main radiative species detected in the He-$N_2$ and He-$O_2$ plasmas operating in SP-DBD configuration. The Figures 4b, 4c, 4d and 4e represent the spectra of OH, $N_2^+$, $N_2$ and O species measured with the 1200 lines.mm$^{-1}$ grating respectively.

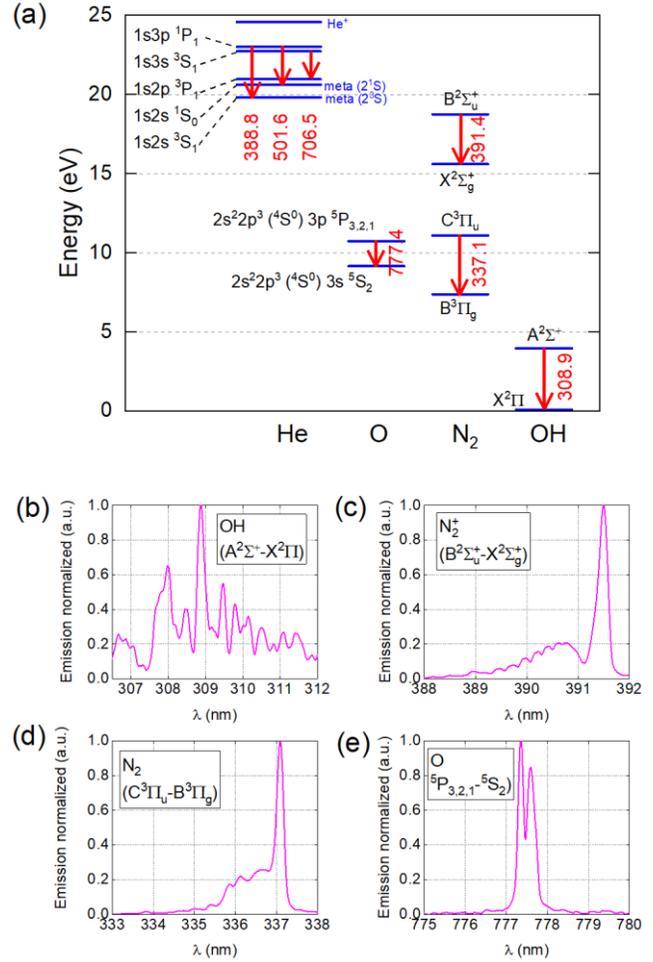

*Figure 4. (a) energetic levels and transitions of the main radiative species detected in He-$N_2$ and He-$O_2$ plasmas in SP-DBD configuration, (b) OH band profile obtained in the He-$N_2$ plasma, (c) $N_2^+$ band profile obtained in the He-$N_2$ plasma, (d) $N_2$ band profile obtained in the He-$N_2$ plasma, (d) O line profile obtained in the He-$O_2$ plasma.*

When a dielectric barrier discharge is only supplied in helium, $He^+$ ions are directly produced by electron impact through {1} (see Table 3) while $He_2^+$ molecular ions result from two-body {2} and three-by collisions {3} with helium species. Then, the production of helium metastable species ($He^m$) can rely on two main reactions which are direct electron impact {4} and dissociative recombination {5}. In turn, the $He(3^3S_1)$ radiative species can be populated following several processes including direct electron impact excitation of helium atoms either in ground state {6} or in metastable state {7} as well as dissociative recombination {8}.

The admixture of molecular nitrogen to the helium gas leads to the formation of molecular ions ($N_2^+$) through several chemical pathways, including charge transfer upon two-body collisions {09} or three-body collisions {10}, but also Penning ionization operating through {11} and {12}. Other reactions including helium and molecular nitrogen species lead to the formation of various products, from {13} to {18}. Another set of reactions occurring only between nitrogen species (between {19} and {22}) is noteworthy, in particular the reaction {19} whose optical transition is clearly observed at 391.4 nm. Finally, since the He-N$_2$ plasma is generated in a dielectric barrier device where a background of ambient air remains present, the presence of water molecules is naturally expected owing to relative humidity. Water protonation {23} and water ionization {24} can be initiated by collisions between helium metastable atoms and nitrogen molecules in the air. As already supported by Xiong et al. [40] that Penning ionization of N$_2$ and subsequent energy transfer to the water molecules are likely to drive to an efficient water ionization and protonation. Then, O and OH radicals could be produced from reaction involving water molecules interplaying either with $He^m$ (reaction {25}), or electrons (reaction {26}) or even $N_2^+$ ions (reaction {27}).

| Reaction | Gas | Rate constant at 300 K | # |
|---|---|---|---|
| $He + e \rightarrow He^+ + 2e$ | He | f(E/n) | {01} |
| $He^*(n>2) + He \rightarrow He_2^+ + e$ | He | f(E/n) | {02} |
| $He^+ + He + He \rightarrow He_2^+ + He$ | He | $1.5 \times 10^{-31} cm^6.s^{-1}$ | {03} |
| $He + e \xrightarrow{E_{thresh}=19.82eV} He^m$ | He | f(E/n) | {04} |
| $He_2^+ + e \rightarrow He^m + He \rightarrow (2He + h\nu)$ | He | $9 \times 10^{-9} cm^3.s^{-1}$ | {05} |
| $He + e \xrightarrow{E_{thresh}=22.72eV} He(3^3S_1) + e$ | He | f(E/n) | {06} |
| $He^m + e \xrightarrow{E_{thresh}=2.9eV} He(3^3S_1) + e$ | He | f(E/n) | {07} |
| $He_2^+ + He \rightarrow He(3^3S_1) + He \rightarrow (2He + h\nu)$ | He | - | {08} |
| $He_2^+ + N_2 \rightarrow N_2^+(B^2\Sigma_u^+) + 2He$ | He-N$_2$ | $8.3 \times 10^{-10} cm^3.s^{-1}$ | {09} |
| $He_2^+ + N_2 + He \rightarrow N_2^+(B^2\Sigma_u^+) + 3He$ | He-N$_2$ | $1.0 \times 10^{-29} cm^3.s^{-1}$ | {10} |
| $He^m + N_2(X^1\Sigma_g^+) \rightarrow N_2^+(B^2\Sigma_u^+) + He + e$ | He-N$_2$ | $7.6 \times 10^{-11} cm^3.s^{-1}$ | {11} |
| $He^m + N_2 + He \rightarrow N_2^+(B^2\Sigma_u^+) + 2He$ | He-N$_2$ | $1.7 \times 10^{-30} cm^6.s^{-1}$ | {12} |
| $He^+ + N_2 \rightarrow products$ | He-N$_2$ | $1.2 \times 10^{-9} cm^3.s^{-1}$ | {13} |
| $He^+ + N_2 + He \rightarrow products$ | He-N$_2$ | $21.2 \times 10^{-29} cm^6.s^{-1}$ | {14} |
| $He_2^+ + N_2 \rightarrow products$ | He-N$_2$ | $2.7 \times 10^{-10} cm^3.s^{-1}$ | {15} |
| $He_2^+ + N_2 + He \rightarrow products$ | He-N$_2$ | $3.4 \times 10^{-30} cm^6.s^{-1}$ | {16} |
| $He^m + N_2 \rightarrow products$ | He-N$_2$ | $1.4 \times 10^{-10} cm^3.s^{-1}$ | {17} |
| $He^m + N_2 + He \rightarrow products$ | He-N$_2$ | $1.7 \times 10^{-30} cm^6.s^{-1}$ | {18} |
| $N_2^+(B^2\Sigma_u^+) \rightarrow N_2^+(X^2\Sigma_g^+) + h\nu$ | He-N$_2$ | $1.5 \times 10^7 s^{-1}$ | {19} |
| $N_2^+(B^2\Sigma_u^+) + e \rightarrow N + N$ | He-N$_2$ | $1.75 \times 10^{-7} cm^3.s^{-1}$ | {20} |
| $N_2^+(B^2\Sigma_u^+) + N_2 \rightarrow products$ | He-N$_2$ | $3.7 \times 10^{-10} cm^3.s^{-1}$ | {21} |
| $N_2(X^1\Sigma_g^+) + e \xrightarrow{E_{thresh}=18.7eV} N_2^+(B^2\Sigma_u^+) + 2e$ | He-N$_2$ | f(E/n) | {22} |
| $H_2O + H^+ + M \rightarrow H^+(H_2O) + M$ | He-H$_2$O | - | {23} |
| $H_2O + M \rightarrow OH^- + H^+ + M$ | He-H$_2$O | - | {24} |
| $He^m + H_2O \rightarrow \begin{cases} OH^*(A^2\Sigma^+) + H + He \\ O^+ + H_2 + He \\ OH(X^2\Pi) + H^* + He \end{cases}$ | He-H$_2$O | - | {25} |
| $H_2O(X^1A_1) + e \xrightarrow{E_{thresh}=10eV} OH(A^2\Sigma)^* + H + e$ | He-H$_2$O | - | {26} |
| $N_2^+(B^2\Sigma_u^+) + H_2O \rightarrow N_2^+(X^2\Sigma_g^+) + OH^* + H$ | He-N$_2$-H$_2$O | - | {27} |
| $N_2(A^3\Sigma_u^+) + OH(X^2\Pi) \rightarrow OH(A^2\Sigma^+) + N_2(X^1\Sigma_g^+)$ | He-N$_2$-H$_2$O | $1.0 \times 10^{-10} cm^3.s^{-1}$ | {28} |
| $O_2 + e \rightarrow O_2^+ + 2e$ | He-O$_2$ | f(E/n) | {29} |
| $O_2^+ + e \rightarrow O + O$ | He-O$_2$ | $4.8 \times 10^{-7} cm^3.s^{-1}$ | {30} |
| $O_2 + e \rightarrow O_2^-$ | He-O$_2$ | f(E/n) | {31} |
| $He^m + O_2 \rightarrow He + O_2^+ + e$ | He-O$_2$ | $2.4 \times 10^{-10} cm^3.s^{-1}$ | {32} |
| $He^m + O \rightarrow He + O^+ + e$ | He-O$_2$ | $4.3 \times 10^{-10} cm^3.s^{-1}$ | {33} |
| $He + O + O \rightarrow He + O_2$ | He-O$_2$ | $1.04 \times 10^{-33} cm^6.s^{-1}$ | {34} |
| $He + O_2 + O \rightarrow He + O_3$ | He-O$_2$ | $6.27 \times 10^{-34} cm^6.s^{-1}$ | {35} |
| $He + O_3 \rightarrow He + O_2 + O$ | He-O$_2$ | $2.28 \times 10^{-26} cm^3.s^{-1}$ | {36} |
| $O_3 + O \rightarrow O_2 + O_2$ | He-O$_2$ | $8.3 \times 10^{-15} cm^3.s^{-1}$ | {37} |
| $O_2 + O^+ \rightarrow O_2^+ + O$ | He-O$_2$ | $2.0 \times 10^{-11} cm^3.s^{-1}$ | {38} |
| $O + O_2 + O_2 \rightarrow O_2 + O_3$ | He-O$_2$ | $6.9 \times 10^{-34} cm^6.s^{-1}$ | {39} |
| $O_2^- + O \rightarrow O_3 + e$ | He-O$_2$ | $1,5 \times 10^{-10} cm^6.s^{-1}$ | {40} |

*Table 3. Reactions and rate constants in He-N2 and He-O2 plasmas at 300 K and atmospheric pressure, considering a residual background of ambient air.*

In Figure 5a, the increase in $\Phi_{N2}$ drives to a slight narrowing of the H$_\alpha$ line FWHM which, in accordance with Stark effect, can be interpreted as a decrease in electron density. Here, electron density is decreased from 5.9 10$^{11}$ cm$^{-3}$ (pure helium) to 3.8 10$^{11}$ cm$^{-3}$ (Helium with 150 sccm of N$_2$). This observation is consistent with the works of Urabe et al. where electron density is estimated to 3.2×10$^{12}$ cm$^{-3}$ for pure helium and 2.3 10$^{12}$ cm$^{-3}$ for 0.5% of nitrogen added [41], as well as with the works of Petrov et al. in the case of He-N$_2$ capillary surface wave discharge [42]. The increase in $\Phi_{N2}$ compel electrons to collide with helium species as well as with an increasing number of molecular nitrogen species. Then, the proportion of electrons involved in reactions {1}, {4} and {5} is lower, hence reducing directly the number of $He^+$ and $He^m$ species and indirectly $He_2^+$ (through {3}) as also suggested by Petrov et al. [42]. As a result, the $He^m$ are fewer to participate in the reactions {11} and {12} as well as {25}, hence explaining the fewer molecular nitrogen ions $N_2^+(B^2\Sigma_u^+)$ produced. Besides, It is worth notifying that the constant rates of these reactions ($7.6 \times 10^{-11} cm^3.s^{-1}$ for the two-body collision {11} and $1.7 \times 10^{-30} cm^6.s^{-1}$ for the three-body collision {12}) are lower than those of the reactions consuming $N_2^+(B^2\Sigma_u^+)$ ($1.5 \times 10^7 s^{-1}$ and $1.75 \times 10^{-7} cm^3.s^{-1}$ for reactions {19} and {20} respectively). This means that $N_2^+(B^2\Sigma_u^+)$ can be consumed at a total rate higher than the one which can produce it and, therefore, why the emission of $N_2^+(B^2\Sigma_u^+)$ is decreased for increasing N$_2$ flow rate. In parallel to a rise in the N$_2$ emission, the Figure 5 shows a significant increase in the optical emission of the OH band. The reaction {25} seems not a relevant pathway to explain this trend since the $He^m$ species are consumed through the aforementioned mechanism. However, direct electron impact as suggested by the reaction {26} seems the most appropriate pathway to explain the production of these hydroxyl radicals.

The admixture of molecular oxygen (O$_2$) to the helium discharge drives to the creation of short and long lifetime radicals and reactive species of oxygen. First, electron collisions lead to the production $O_2^+$ through reaction {29} and subsequently to O radicals {30}. Depending on the value of the reduced electric field (E/n), $O_2^-$ (superoxide anion) can also be produced according {31}. In addition, the $He^m$ species (whether $He(2^1S)$ or $He(2^3S)$) contribute to the formation of $O^+$ and $O_2^+$ ions through Penning ionization reactions {32} and {33} respectively. $O_3$ can also be created through three-body collisions involving He, O and O$_2$ {35} although this reaction is counter-balanced by reaction {36} with higher rate constant. Finally, all these oxygenated species can interplay with each other and level off their respective populations (reactions from {37} to {40}).

In Figure 5b, increasing values of $\Phi_{O2}$ drive to a slight narrowing of the H$_\alpha$ line associated with a decrease in the electron density. Values close to 5.9 10$^{11}$ cm$^{-3}$ are estimated in the pure helium case while a gradual decrease is obtained with values as low as 2.1 10$^{10}$ cm$^{-3}$ when 150 sccm of O$_2$ is admixed with helium. Such trend is consistent with the works of Xiao et al. [54]. The reaction {31} can be considered as responsible for such trend since the superoxide anion results from the addition of an electron which fills one of the two degenerate molecular orbitals of O$_2$, leaving a charged ionic species with a single unpaired electron and a net negative charge of −1. Since electron density decreases for increasing values of $\Phi_{O2}$, the density of $He^m$ species decreases too, as well as the intensities of O and OH radicals. The case of $O_3$ is different in that an increase in $\Phi_{O2}$ contributes to strengthen the generation of ozone mainly through reactions {35}, {39} and {40}. Additional reactions can also be involved, as detailed in [51], [52] and [53].

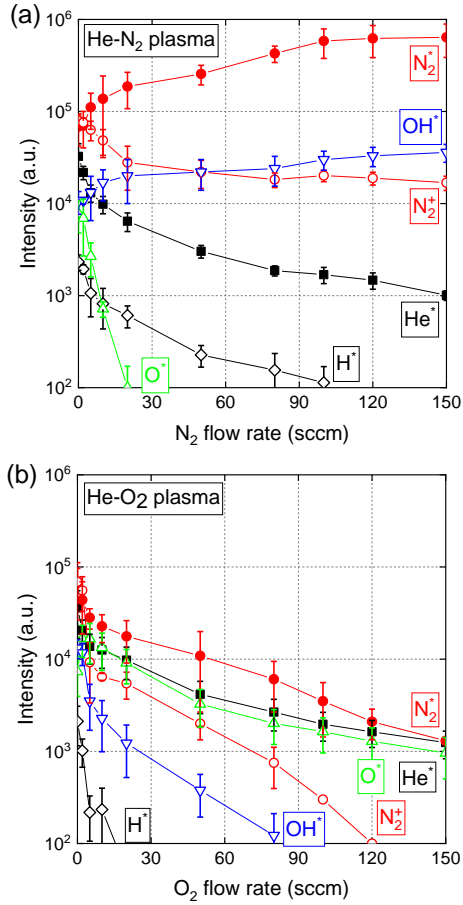

*Figure 5. Optical emission of plasma radiative species as a function of (a) $N_2$ flow rate or (b) $O_2$ flow rate. In all cases, the reactive gas ($N_2$ or $O_2$) is mixed with 2 slm of helium; all measurements are achieved in SP-DBD configuration.*

### III.2. Influence of plasma exposure time

The Figure 6a shows the germination curves of seeds without treatment (control group, black square symbols) and seeds exposed to a pure helium plasma (plasma group, open circle symbols). This treatment has been achieved using helium gas at 2 slm without any reactive gas admixture. Each data point corresponds to an average value resulting from 5 samples, each sample containing 80 seeds. If the plasma treatment does not change the germination rate ($G_\%^{max}$) which is already closed to 100% or the homogeneity, it has however a strong impact on the vigor. Indeed, the $\tau_{50}$ indicator decreases from 1850 min to 1610 min, which represents a time saving of approximately 4 hours ($\Delta\tau = 240$ min), a relative gain of $\frac{240}{1850} \times 100 = 12.9\%$.

To investigate to which extent the plasma exposure time can reduce $\tau_{50}$, the SP-DBD has been operated from 1 min to 90 min under the same experimental conditions (2 slm of helium). The Figure 6b shows a non-linear relation between exposure time and $\tau_{50}$. Even if $\tau_{50}$ values as low as 1470 min can be reached after 1h30 of plasma treatment, 90% of the $\tau_{50}$ variations occur after 10 min of exposure. For all the experiments carried out hereafter, the plasma exposure times have been fixed at 20 min.

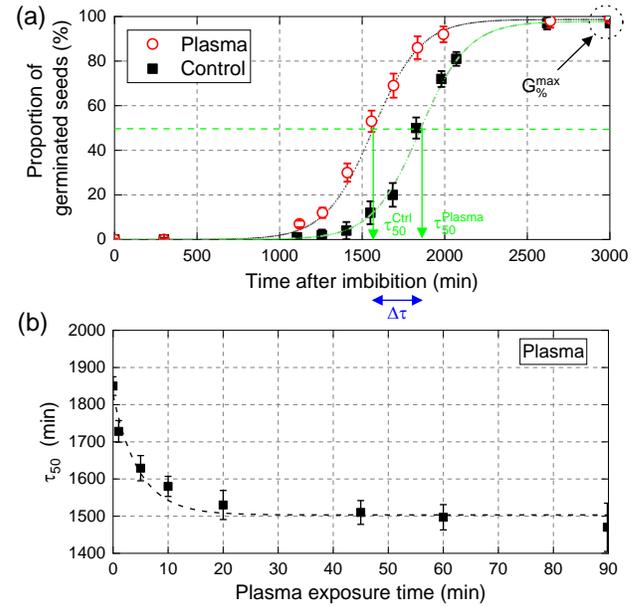

*Figure 6 (a) Germination curves of lentils either untreated or treated with plasma (pure Helium at 2slm, 6kV, 600Hz), (b) Influence of the plasma exposure time on the $\tau_{50}$ indicator with same experimental conditions as (a). Values are means ± SE (n = 5, 5×80 = 400 seeds).*

### III.3. Influence of plasma gaseous chemistry on seeds vigor and related ageing effects

To verify if plasma reactive chemistry can promote seeds biological properties, plasma has been generated during 20 min using a mixture of helium (carrier gas) with molecular nitrogen or oxygen. The helium flow rate has been set at 2 slm while the reactive gas flow rate has been varied between 0 and 150 sccm. The Figure 7a shows the effect of the two plasma treatments on the vigor of lentil seeds for increasing values of $\Phi_{N2}$ (black filled square symbols) or of $\Phi_{O2}$ (black open square symbols).

If a time saving as high as 240 min is obtained by supplying plasma only with helium, it turns out that the admixture of 150 sccm of nitrogen leads to $\Delta\tau$ close to 480 min (8 hours) which represents a relative gain of 480/1850 = 25.9%. This latter value has to be compared with the gain of a pure He plasma treatment which is 12.9%). On the other hand, the addition of $O_2$ slightly slows down the $\Delta\tau$ values, decreasing to 180 min for $\Phi_{O2}$ = 150 sccm. These results show that the plasma reactive chemistry can modulate seeds vigor, either through stimulating effects ($N_2$ plasma) or inhibitory effects ($O_2$ plasma).

To check whether the time saving ($\Delta\tau$) remains sustainable over ageing, a large amount of lentil seeds has been exposed to plasma and stored in dark (T=20°C, RH=40%) during 15, 30 or 45 days, before being soaked with tap water for germination follow up. As shown in Figure 7a, whatever the DAPP (He-$N_2$ or He-$O_2$), no strong difference appears compared with the curves at 0 day (no ageing).

The Figure 7b offers a more explicit representation of $\Delta\tau$'s sustainability over 45 days of aging by focusing on three plasma treatments (with $\Phi_{He}$ =2 slm, 20 min) : pure He treatment, He-N$_2$ treatment with $\Phi_{N_2}$ = 150 sccm and He-O$_2$ treatment with $\Phi_{O_2}$ =150 sccm. Overall, it turns out that $\Delta\tau$ slightly decreases the first 15 days and that the subsequent variations remain poorly significant. As an example, in the case of the pure He treatment, $\Delta\tau$ decreases from 284 min to 257 min. The indicators of homogeneity and germination rate have also been measured but, as reported in Table 4, no statistical modifications could be evidenced, whatever the reactive gas flow rates.

| $\Phi$ (sccm) | Germination slope indicator ($S$) | | Germination rate indicator ($G_\%^{max}$) | |
|---|---|---|---|---|
| | He-N$_2$ plasma | He-O$_2$ plasma | He-N$_2$ plasma | He-O$_2$ plasma |
| 0 | 39.7 (±1.7) | | 98.5 (±0.3) | |
| 2 | 33.3 (±3.7) | 32.8 (±6.7) | 99.3 (±1.4) | 98.7 (±0.7) |
| 20 | 38.6 (±1.6) | 32.5 (±1.4) | 99.2 (±0.8) | 99.3 (±0.6) |
| 50 | 32.8 (±1.9) | 36.4 (±4.5) | 98.5 (±0.6) | 97.6 (±0.8) |
| 80 | 34.5 (±2.7) | 32.5 (±1.6) | 99.3 (±1.3) | 98.8 (±0.5) |
| 100 | 37.2 (±5.8) | 37.2 (±4.1) | 99.5 (±0.9) | 98.9 (±1.6) |
| 120 | 35.8 (±4.3) | 31.5 (±2.8) | 97.0 (±0.9) | 99.4 (±0.3) |
| 150 | 39.1 (±4.2) | 38.9 (±3.1) | 100.0 (±0.8) | 97.6 (±0.6) |

Table 4. Influence of He-N$_2$ and He-O$_2$ plasma treatments on the germination slope and germination rate indicators standing for homogeneity and germination potential respectively. Conditions: treatment time=20min, He gas flow rate = 2 slm, f=600 Hz, Amplitude= 6 kV. Values are means ± SE (n = 5, 5×80 = 400 seeds).

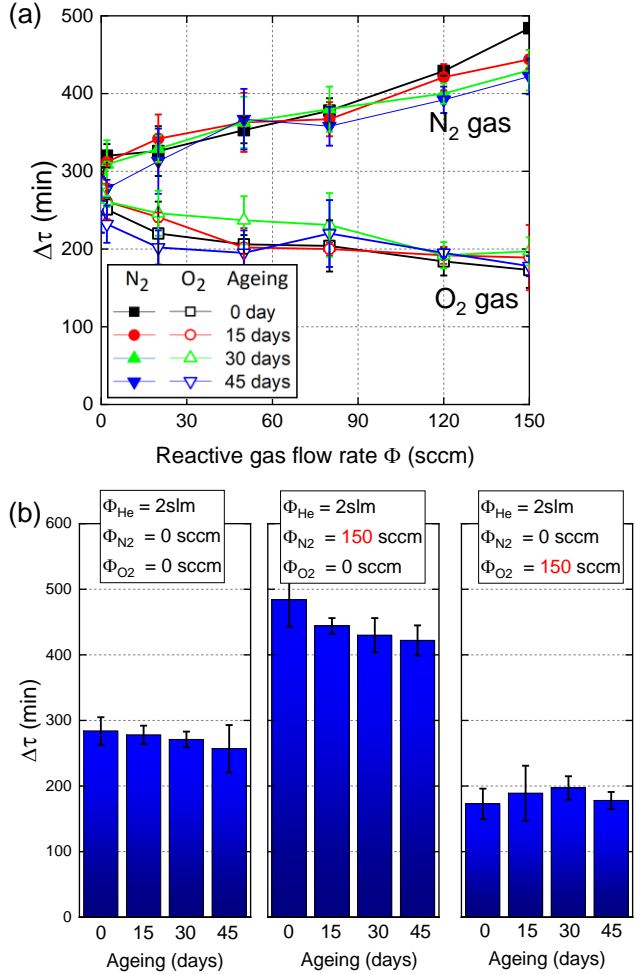

To understand which active species from the plasma phase are likely to generate a sustainable vigor, a correlation diagram is plotted in Figure 8 where (i) the $\Delta\tau$ values are represented by colored circles and (ii) the intensities of N$_2$* N$_2^+$, NO, O, O$_3$ and OH species previously identified by mass spectrometry and optical emission spectroscopy are normalized between 0 and 1. Since they cannot produce one single type of active species, the atmospheric He-N$_2$ and HeO$_2$ plasmas cannot be considered as highly chemo-selective, leading us to only state that N$_2$*, N$_2^+$, OH and NO together promote seed vigor in a significant way without pointing out which of these species plays a leading role. On the contrary, we can infer that O species have limited beneficial effects on $\Delta\tau$ while O$_3$ clearly shows inhibitory effects on seeds vigor.

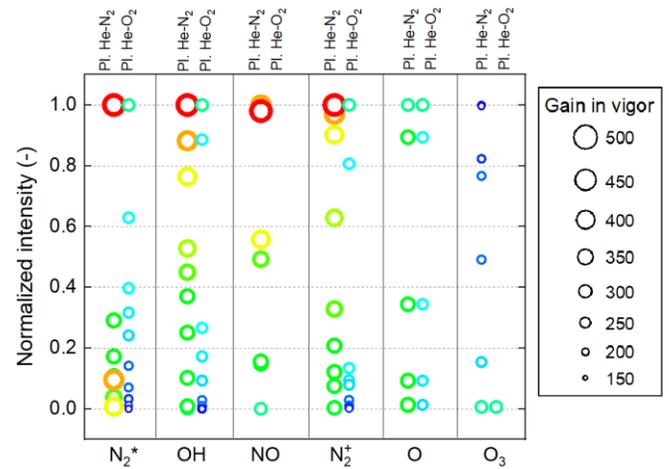

Figure 8. Correlation diagram between the gain in vigor of lentil seeds ($\Delta\tau$) with normalized intensities of active species measured by mass spectrometry and optical emission spectroscopy in the plasma phase.

Figure 7. (a) Variation of the $\Delta\tau$ indicator as a function of the reactive gas flow rate (nitrogen or oxygen) admixed with helium (2 slm) in the SP-DBD. Seeds are imbibed with water either at 0, 15, 30 or 45 days after plasma treatment (b) Variations of the gain in vigor as a function of ageing time for 3 plasma treatments of 20 min: pure He, He-N$_2$/150sccm and He-O$_2$/150 sccm. Values are means ± SE (n = 5, 5×80 = 400 seeds). All the p-values on the $\tau_{50}$ parameter are lower than 0.05.

### III.4. Influence of seeds coating wettability and seed water uptake

It is questionable whether the plasma-improved seed vigor is due to better interfacing with water, either through modifications of its surface wettability properties and/or its water uptake capacity. In addition to wondering whether these modifications are significant, the question of their resistance to time is raised: long-term effects similar to the one obtained on vigor (45 days) are expected.

### III.4.1. DAPP increases seeds coating wettability

The wettability state of any surface, its hydrophilic/hydrophobic nature, is the result of surface physical properties (roughness, multi-scale patterns) as well as chemical properties (functionalization, oxidation). To assess the wettability state of seeds teguments before/after plasma exposure, water contact angles have been measured by depositing a drop of milli-Q water on their surface. The Figure 9 shows some photographs of the seeds, including untreated seeds (control) with an average WCA value as high as 117.8°, seeds exposed to pure helium with WCA=38.9°, seeds exposed to He-$O_2$ plasma with WCA=24.7°, etc. The influence of increasing $N_2$ and $O_2$ flow rates on seeds teguments is reported in Figure 10a and 10b respectively for different ageing times. When seeds are immediately imbibed after plasma exposure ($t_{ageing}$ = 0 day), the values decrease from 35° (pure He) to approximately 25° for 150 sccm of molecular nitrogen or oxygen. The admixture of a reactive gas to helium contributes to the hydrophilization of the coatings, at least on the topmost surface layer (typically < 100 nm indepth). To verify if such plasma-driven hydrophilic state is sensitive to ageing effects, we have proceeded to WCA measurements 15, 30 and 45 days after their plasma exposures. Although the native WCA value is never reached even 45 days later, a partial recovery of seeds wettability surface is obtained, with WCA values increasing up to 60° and 37° for 150 sccm of nitrogen and oxygen respectively. In overall, this WCA recovery is stronger with the He-$N_2$ plasma rather than with the He-$O_2$ plasma.

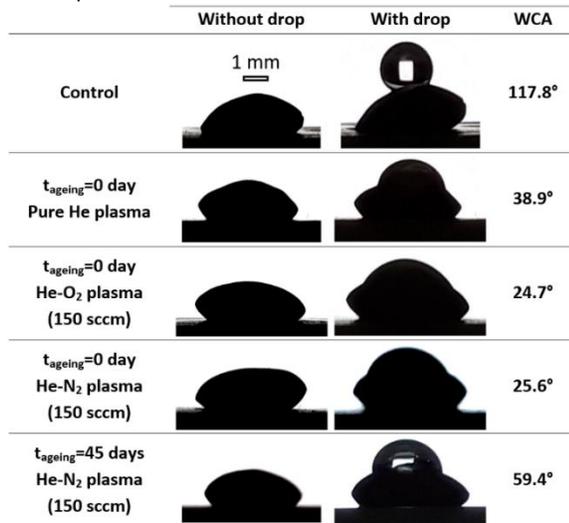

*Figure 9. Photographic views of lentils seeds without/with drops (milli-Q water) deposited and analyzed following the Sessile drop method.*

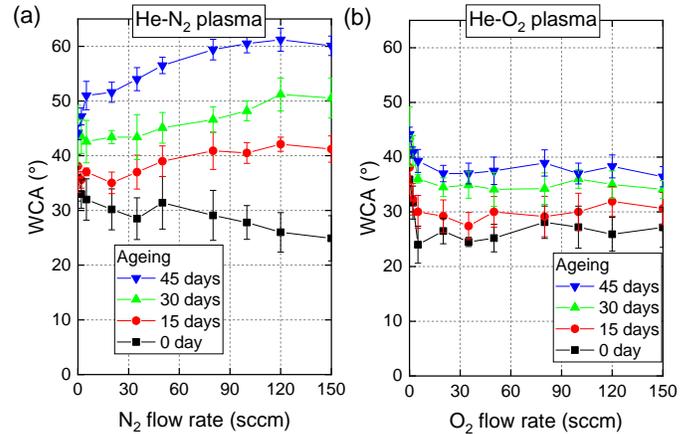

*Figure 10. Water contact angles measured on seeds lentils after (a) He-$N_2$ or (b) He-$O_2$ plasma treatments. Measurements were achieved with the drop shape analyzer at $t_{ageing}$ = 0, 15, 30 or 45 days after plasma treatment. Each data point is the statistical average of 10 WCA achieved on 10 distinct seeds.*

### III.4.2. DAPP increases water uptake

Since plasma treatments can durably change seeds coating wettability, one may wonder whether plasma can modify the water absorption properties inside the seeds. The Figure 11a reports a series of photographs of the same lentil seed taken at different time intervals after its imbibition in water (t = 0h). Since a clear and progressive swelling of the lentil is observed after plasma exposure, measuring water uptakes as defined in equation (2) appears appropriate. This monitoring has been carried out during 2h, 4h, 6h and 8h after imbibition: a period along which the stronger variations in water uptake are observed and correspond to protein activation, protein synthesis and ignition of seeds respiration.

The Figure 11b shows water uptakes as a function of reactive gas flow rate ($N_2$ or $O_2$) considering the seeds 2 hours and 8 hours after their imbibition. First, it turns out that exposing lentil seeds to a pure helium plasma significantly increases their $\xi$. Two hours after imbibition, $\xi$ goes from 10% (control) to 15% ($\Phi$ = 0sccm). Then, this increase can be further improved by admixing a small quantity of reactive gas to the helium: 2 hours after imbibition, $\xi$ rises from 15% (pure He) to 23% (He with reactive gas). It should be emphasized that the nature of the reactive gas, whether $O_2$ or $N_2$, leads to the same values of $\xi$. This trend is also confirmed for measurements of water uptakes carried out 8 hours after imbibition with values increasing from 42% ($\Phi$ = 0 sccm) to approximately 52% ($\Phi$ = 150 sccm) whatever the reactive gas. In addition, Figure 11c shows water uptake gains which are durable, at least over an aging period of 45 days: for a flow rate of 150 sccm and a post-imbibition time of 8 hours, the $\xi$ values turn around 50% (instead of 52% in Figure 11b).

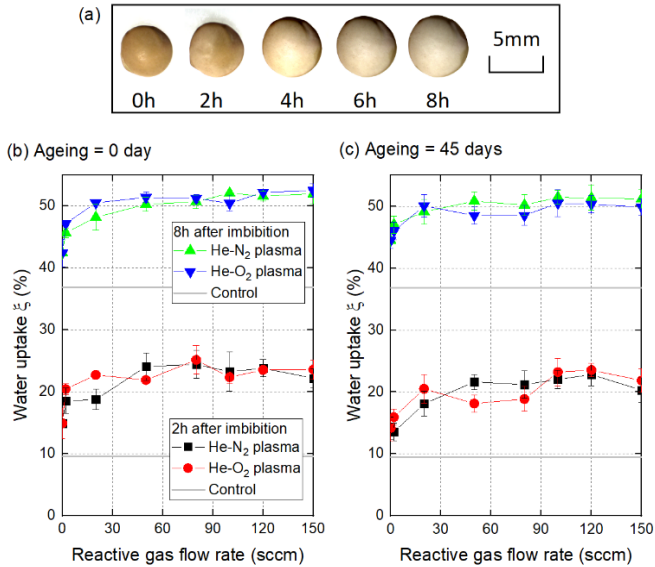

*Figure 11. (a) Time follow-up of lentil seeds after imbibition with water, (b, c) Water uptake of lentil seeds exposed to He-N$_2$ or He-O$_2$ plasmas as a function of reactive gas flow rate admixed with helium gas (2 slm). Seeds imbibition is achieved either (b) immediately after plasma treatment or (c) 45 days after.*

## III.5. Inhibitory and stimulating effects of the reactive species mediated by the seed's water circulation system

If DAPP can hydrophilize seeds surface and increase their ability in absorbing more water, it is questionable if these two modifications contribute to improve the vigor. To answer this question, the WCA and $\xi$ parameters are plotted as a function of $\Delta\tau$ in Figures 12a and 12b respectively while considering several ageing times. These figures show that upon their ageing, the plasma-activated seeds exhibit a low loss of their vigor during the first two weeks and stabilize thereafter. This loss of vigor is accompanied by significant changes in the WCA values while the water uptake values remain almost unchanged over a 45 days ageing period.

The WCA measurements evidence that DAPP induces a metastable chemical functionalization, with probably the grafting of polar groups containing oxygen and/or nitrogen like hydroxyl groups, carbonyl groups or carboxylic acids as often observed in other applications, e.g. carbon scaffolds, polymers exposed to cold atmospheric plasmas [55]. Besides, these WCA analyzes do not necessarily mean that the (super) hydrophilic character of the seed surface is responsible for improving vigor or that the partial recovery state after 45 days is responsible for a decay of $\Delta\tau$. Indeed, these WCA measurements could also be interpreted as a proxy picture of deeper modifications. One must keep in mind that seeds are porous biological structures and that assessing the penetration depth of reactive species would deserve additional investigations. For this reason, we assume that the plasma-induced functionalization does not only operate on the seed outer surface but could also cover deeper biological tissues. With ageing, the reactive species grafted on the topmost outer layers of the tegument are expected to gradually disappear, their lifespan being on the order of a few weeks.

Conversely, the plasma-treated seeds show water uptake values which remain both elevated and stable over the 45 days ageing period. These plasma-induced irreversible modifications may result from two mechanisms: (i) a chemical mechanism governed by the in-depth grafting of hydrophilic functions, protected from the environment and exhibiting lifespan higher than 45 days and/or (ii) a physical mechanism where plasma could create stable structural modifications inside the seed through cleaning, widening, smoothing, etching, remodeling pores and therefore redefining water imbibition efficiency.

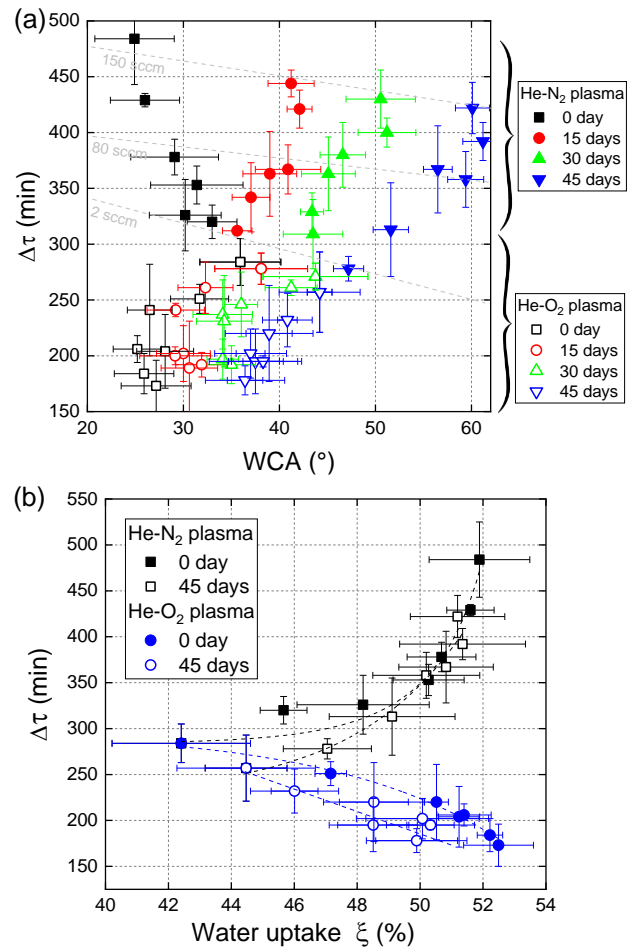

*Figure 12. Correlation diagram between (a) $\Delta\tau$ and WCA measurements, (b) $\Delta\tau$ and water uptakes measurements. In both cases, He-N$_2$ or He-O$_2$ plasma treatments are considered for several ageing times (0, 15, 30 and 45 days). Each data point is the statistical average of 5×80 = 400 seeds.*

Overall, it turns out that the WCA values as well as the water uptakes are not obviously correlated with the sustainable increases in vigor. As an example, the Figure 12a shows that for a same WCA value of 40°, the gain in vigor can change over a wide range of values (between 170 and 480 min) while for $\xi$=52%, the Figure 12b indicates a lower $\Delta\tau$ value (170 min) with the He-O$_2$ plasma treatment vs a higher one (480 min) with the He-N$_2$ plasma

*This document is a pre-print version. You may use it at your own convenience although its content may deviate in places from the final published article.*treatment. Considering these results as well as those from Figure 8, we assume that DAPP improves the efficiency of the seed hydric circulation system, from the outer surface to the inner tissues of the seeds. This claim is consistent with the works of P. Cloetens et al. where, using quantitative phase tomography with synchrotron radiation, the authors have shown the existence of a 3D network of intercellular air space in Arabidopsis seeds, likely to serve for rapid water uptake and distribution during imbibition [56]. In our case, plasma is assumed to enhance this inner water transport network, carrying more efficiently the reactive species whether plasma-grafted or plasma-triggered (i.e. produced by the seeds in reaction to the plasma species). Whatever their origin, these reactive species can have stimulating or inhibitory effects on $\Delta\tau$. In the case of a He-N$_2$ plasma, active species like N$_2$* and OH can efficiently promote the seed vigor while in the case of He-O$_2$ plasma, O$_3$ can counter-balance the stimulating effects of the other active species, hence reducing the time saving.

## IV. CONCLUSION

Lentil seeds have been treated using a SP-DBD which is one device among others to perform a dry atmospheric plasma priming (DAPP). Whether supplied in helium gas with/without admixture of a reactive gas, no effect has been evidenced on germination rate and homogeneity of lentil seeds. However, important effects have been demonstrated on their vigor. While a pure helium plasma improves this germination indicator by approximately 5 hours compared with control, the admixture of molecular nitrogen increases this rate up to 8 hours. The active species from the plasma assumed to sustain this biological effect are N$_2$*, N$_2^+$, OH and NO. By diffusing inside the seeds, they are expected to form long lifespan reactive species which remain grafted within seeds inner tissues. Indirect evidence corroborate this claim, first the super-hydrophilic state of seeds surface with WCA values as low as 25° (instead of native WCA of 118°) and second higher water uptakes (as high as 50% after a post-imbibition delay of 8 hours). In the case of a He-O$_2$ plasma, the actives species are expected to induce competitive effects, in particular with the production of ozone which induces inhibitory effect on vigor magnitude. The plasma active species likely to diffuse into the seed and modify its vigor are not necessarily the same as those involved in the modifications of surface WCA and water uptakes. DAPP is assumed to reinforce seeds hydric circulation system, facilitating the interplay between the reactive species and seed's inner tissues. Whether these reactive species are plasma-grafted or plasma-triggered, their stimulating or inhibitory effects are mediated through this enhanced tridimensional network.

The ageing of the seeds 15, 30 and 45 days after DAPP shows that the gain in vigor remains sustainable over more than one month, even if a slight decrease is observed, e.g. 30 minutes in the case of the pure He plasma treatment. Seeking optimal storage conditions (temperature, humidity) could offer some leeway to further improve the sustainability of $\Delta\tau$, i.e. larger values that remain constant over longer periods. These long-term effects have also been demonstrated on the water uptakes and WCA values measured along 45 days.

This research work opens the way to future thorough investigations where the influence of seeds rubbery and glassy states (that depend on their water content and ambient temperature) will be considered, as well as modifications in hormonal activities.

## V. ACKNOWLEDGEMENTS

This work has been supported by grants from Région Ile-de-France (Sesame, Ref. 16016309) and ABIOMEDE platform through the Sorbonne Université Platform programme. This work is partly supported by French network GDR 2025 HAPPYBIO.## VI. DATA AVAILABILITY STATEMENT

The data that support the findings of this study are available from the corresponding author upon reasonable request.

## VII. REFERENCES

[1] Seed technology and its biological basis, M. Black, J. Derek Bewley, 2000, ISBN 1-84127-043-1

[2] Influence of priming techniques on seed germination behavior of maize inbred lines (Zea mays L.), P. M. Dezfuli, F. Sharif-Zadeh, M. Janmohammadi, ARPN Journal of Agricultural and Biological Science, Vol. 3, No. 3, ISSN 1990-6145, (2008)

[3] Effects of different priming techniques on seed invigoration and seedling establishment of lentil (Lens culinaris Medik), K. Ghassemi-Golezani, A. A. Aliloo, M. Valizadeh, M. Moghaddam, Journal of Food, Agriculture & Environment, Vol. 6, No. 2, 222-226 (2008)

[4] Seed priming improves the germination and growth rate of melon seedlings under saline stress, C. Eduardo da Silva Oliveira, F. Steiner, A. M. Zuffo, T. Zoz, C. Z. Alves, V. Cabrera, B. de Aguiar, Ciência Rural, Santa Maria, v.49:07, e20180588, ISSNe 1678-4596 (2019)

[5] Inducing rapid seed germination of native cool season grasses with solid matrix priming and seed extrusion technology, M. D. Madsen, L. Svejcar, J. Radke, A. Hulet, PLoSONE 13(10):e0204380 (2018)

[6] Performance of bell pepper seeds in response to drum priming with addition of 24-epibrassinolide, C. Barboza da Silva, J. Marcos-Filho, P. Jourdan, M. A. Bennett, HortScience, 50(6):873-878 (2015)

[7] Surface chemistry and germination improvement of Quinoa seeds subjected to plasma activation, A. Gómez-Ramírez, C. López-Santos, M. Cantos, J. L. García, R. Molina, J. Cotrino, J. P. Espinós, A. R. González-Elipe, Scientific Reports, Vol. 7, 5924 (2017)

[8] Cold radiofrequency plasma treatment modifies wettability and germination speed of plant seeds, E. Bormashenko, R. Grynyov, Y. Bormashenko, E. Drori, Scientific reports, 2:741 (2012)

[9] The stimulatory effect of non-equilibrium (low temperature) air plasma pretreatment on light-induced germination of *Paulownia tomentosa*
13*T. Dufour et al.*

seeds, S. Zivkovic, N. Puac, Z. Giba, D. Grubisic, Z. L. J. Petrovic, SeedSci. & Technol., 32, 693-701 (2004)

[10] Using low-pressure plasma for Carthamus tinctorium L. seed surface modification, M. Dhayala, S.-Y. Lee, S.-U. Park, Vacuum, Vol. 80, Issue 5, 499-506 (2006)

[11] Effects of high voltage nanosecond pulsed plasma and micro DBD plasma on seed germination, growth development and physiological activities in spinach, S.-H. Ji, K.-H. Choi, A. Pengkita, J. S. Im, J. S. Kim, Y. H. Kim, Y. Park, E. J. Hong, S. K. Jung, E.-H. Choi, G. Park, Archives of Biochemistry and Biophysics, Vol. 605, 117-128 (2016)

[12] Effect of low-temperature plasma on the structure of seeds, growth and metabolism of endogenous phytohormones in pea (Pisum sativum L.), T. Stolárik, M. Henselová, M. Martinka, O. Novák, A. Zahoranová, M. Černák, Plasma Chemistry and Plasma Processing, Vol. 35, 659-676 (2015)

[13] Interaction of cold radiofrequency plasma with seeds of beans (Phaseolus vulgaris), E. Bormashenko, Y. Shapira, R. Grynyov, G. Whyman, Y. Bormashenko, E. Drori, Journal of Experimental Botany, Vol. 66, No. 13, pp. 4013-4021 (2015)

[14] Non-Thermal Plasma Treatment of Agricultural Seeds for Stimulation of Germination, Removal of Surface Contamination and Other Benefits: A Review, L. K. Randeniya, G. J. J. B. de Groot, Plasma processes and polymers, Vol. 12, Issue7, 608-623 (2015)

[15] New insights of low-temperature plasma effects on germination of three genotypes of Arabidopsisthaliana seeds under osmotic andsaline stresses, M. Bafoil, A. Le Ru, N. Merbahi, O. Eichwald, C. Dunand, M. Yousfi, Scientific reports, 4:5859 (2014)

[16] Modification of Seed Germination Performance through Cold Plasma Chemistry Technology, J. C. Volin, F. S. Denes, R. A. Young, S. M. T. Park, Crop Science, Issue6, 1706-1718 (2000)

[17] Effect of seed treatment by cold plasma on the resistance of tomato to ralstonia solanacearum (Bacterial Wilt), J. Jiang, Y. Lu, J. Li, L. Li, X. He, H. Shao,Y. Dong, PLoS ONE 9(5): e97753 (2014)

[18] Germination of chenopodium album in response to microwave plasma treatment, B. Sera, V. Stranak, M. Sery, M. Tichy, P. Spatenka, Plasma Science and Technology, Vol.10, No.4 (2008)

[19] Effect of glow discharge air plasma on grain crops seed, A.E. Dubinov, N. Novgorod, E. R. Lazarenko ; V. D. Selemir, IEEE Transactions on Plasma Science, Vol. 28 , Issue: 1 (2000)

[20] Growth, anatomy and enzyme activity changes in maize roots induced by treatment of seeds with low-temperature plasma, M. Henselová, L. Slováková, M. Martinka, A. Zahoranová, Biologia, Section Botany, 67/3: 490-497 (2012)

[21] Positive effect of non-thermal plasma treatment on radish seeds, A. L. Mihai, D. Dobrin, M. Magureanu, M. E. Popa, Romanian Reports in Physics, Vol. 66, No. 4, 1110-1117 (2014)

[22] Non-thermal plasma for germination enhancement of radish seeds, K. Matra, Procedia Computer Science, 86, 132-135 (2016)

[23] Stimulating effects of seed treatment by magnetized plasma on tomato growth and yield, Y. Meiqianf, H. Mingjing, M. Buzhou, M. Tengcai, Plasma Science & Technology, vo1.7, No.6 (2005)

[24] Stimulation of the germination and early growth of tomato seeds by non-thermal plasma, M. Măgureanu, R. Sîrbu, D. Dobrin, M. Gîdea, Plasma Chemistry and Plasma Processing, Vol. 38, 989-1001 (2018)

[25] Effect of Cold Plasma Treatment on Seed Germination and Growth of Wheat, J. Jiang, X. He, L. Li, J. Li, H. Shao, Q. Xu, R. Ye, Y. Dong, Plasma Science and Technology, Vol.16, No.1 (2014)

[26] Influence of plasma treatment on wheat and oat germination and early growth, B. Sera, P. Spatenka, M. Sery, N. Vrchotova, I. Hruskova, IEEE Transactions on plasma science, Vol. 38, No 10 (2010)

[27] The effect of non-thermal plasma treatment on wheat germination and early growth, D. Dobrin, M. Magureanu, N. Bogdan Mandache, M.-D. Ionita, Innovative Food Science & Emerging Technologies, Vol. 29, 255-260 (2015)

[28] Cold plasma treatment enhances oilseed rape seed germination under drought stress, Li Ling, Li Jiangang, Shen Minchong, Zhang Chunlei, Dong Yuanhua, Scientific Reports, Vol. 5, 13033 (2015)

[29] An improved process for high nutrition of germinated brown rice production: Low-pressure plasma, H. H. Chen, H. C. Chang, Y. K. Chen, C. L. Hung, S. Y. Lin, Y. S. Chen, Food Chemistry, Vol. 191, 120-127 (2016)

[30] Effects of cold plasma treatment on seed germination and seedling growth of soybean, L. Ling, J. Jiafeng, L. Jiangang, S. Minchong, H. Xin, Shao Hanliang, D. Yuanhua, Scientific reports, 4:5859 (2014)

[31] Effect of glow discharge plasma on germination and fungal load of some cereal seeds, M. Brasoveanu, M. R. Nemtanu, C. Surdu-Bob, G. Karaca, I. Erper, Romanian Reports in Physics, Vol. 67, No. 2, 617-624 (2015)

[32] Decontamination of grains and legumes infected with Aspergillus spp. and Penicillum spp. by cold plasma treatment, M. Selcuk, L. Oksuz, P. Basaran, Bioresource Technology, Vol. 99, Issue 11, 5104-5109 (2008)

[33] Inactivation of surface-borne microorganisms and increased germination of seed specimen by cold atmospheric plasma, A. Mitra, Y.-F. Li, T. G. Klämpfl, T. Shimizu, J. Jeon, G. E. Morfill, J. L. Zimmermann, Food Bioprocess Technol., 7:645-653 (2014)

[34] Plasma inactivation of microorganisms on sprout seeds in a dielectric barrier discharge, D. Butscher, Hanne Van Loon, A. Waskow, P. Rudolf von Rohr, M. Schuppler, International Journal of Food Microbiology, Vol. 238, 222-232 (2016)

[35] Effects of low-temperature plasmas and plasma activated waters on Arabidopsis thaliana germination and growth, M. Bafoil, A. Jemmat, Y. Martniez, N. Merbahi, O. Eichwald, C. Dunand, M. Yousfi, Plos ONE 13(4): e0195512

[36] Global lentil production :constraints and strategies, S. Kumar, S. Barpete, J. Kumar, P. Gupta, A. Sarker, SATSA Mukhapatra - Annual Technical Issue 17, ISSN 0971-975X (2013)

[37] Seed Germination: mathematical representation and parameters extraction, Y. A. El-Kassaby, I. Moss, D. Kolotelo, M. Stoehr, Forest Science 54(2), 220-227 (2008)

[38] A guide to forest seed handling, R. L. Willan, FAO Forestry paper, Chapter 9, 20/2, M-31, ISBN 92-5-102291-7 (1987), http://www.fao.org/3/AD232E/AD232E00.htm#TOC

[39] http://www.eaudeparis.fr/la-qualite-de-leau-a-paris/

[40] Temporal and spatial resolved optical emission behaviors ofa cold atmospheric pressure plasma jet, Q. Xiong, X. Lu, J. Liu, Y. Xian, Z. Xiong, F. Zou, C. Zou, W. Gong, J. Hu, K. Chen, X.Pei, Z. Jiang, Y. Pan, J. Appl. Phys. 106 (8):083302 - 083302-6 (2009)

[41] Combined spectroscopic methods for electron-density diagnostics inside atmospheric-pressure glow discharge using He/N2 gas mixture, K. Urabe, O. Sakai, K. Tachibana, Journal of Physics D: Applied Physics, Vol. 44, No. 11 (2011)

[42] Numerical modeling of a He-N$_2$ capillary surface wave discharge at atmospheric pressure, G. M. Petrov, J. P. Matte, I. Peres, J. Margot, T. Sadi, J. Hubert, K. C. Tran, L. L. Alves, J. Loureiro, C. M. Ferreira, V. Guerra, G. Gousset, Plasma Chemistry and Plasma Processing, Vol. 20, No. 2 (2000)

[43] Numerical modeling of the effect of the level of nitrogen impurities in a helium parallel plate dielectric barrier discharge, C. Lazarou, D. Koukounis, A. S. Chiper, C. Costin, I. Topala, G. E. Georghiou, Plasma Sources Science and Technology, Vol. 24, No 3 (2015)

[44] Characterization of a dielectric barrier discharge operating in an open reactor with flowing helium, G. Nersisyan, W. G. Graham, Plasma Sources Sci. Technol., Vol. 13, 582-587 (2004)

[45] Variations of the gas temperature in He/N2 barrier discharges, N. K. Bibinov, A. A. Fateev, K. Wiesemann, Plasma Sources Science and Technology, Vol. 10, No 4 (2001)

[46] Spatio-temporally resolved spectroscopic diagnostics of the barrier discharge in air at atmospheric pressure, K. V. Kozlov, H.-E. Wagner, R. Brandenburg, P. Michel, Journal of Physics D: Applied Physics, Vol. 34, No 21 (2001)

[47] Atmospheric pressure plasma jets generated by the DBD in argon-air, helium-air, and helium-water vapour mixtures, A. A. Heneral, S. V. Avtaeva, Journal of Physics D: Applied Physics, Vol. 53, No 19 (2020)

[48] Operation modes of the helium dielectric barrier discharge for soft ionization, S.Müller, T. Krähling, D. Veza, V. Horvatic, C. Vadla, J. Franzke, Spectrochimica Acta Part B: Atomic Spectroscopy, Vol. 85, 104-111 (2013)

[49] Measurement of OH radicals at state X 2Π in an atmospheric-pressure micro-flow dc plasma with liquid electrodes in He, Ar and N2 by means of laser-induced fluorescence spectroscopy, L. Li, A. Nikiforov, Q. Xiong, X. Lu, L. Taghizadeh, C. Leys, Journal of Physics D: Applied Physics, Vol. 45, No. 12 (2012)

[50] On the influence of metastable reactions on rotational temperatures in dielectric barrier discharges in He-N2 mixtures, N. K. Bibinov, A. A. Fateev, K. Wiesemann, Journal of Physics D: Applied Physics, Vol. 34, No. 12 (2001)

[51] Numerical simulation on mode transition of atmospheric dielectric barrier discharge in helium–oxygen mixture, D. Lee, J. M. Park, S. H. Hong, Y. Kim, IEEE Transactions on plasma science, Vol. 33, No. 2, 949-957 (2005)

[52] Formation of reactive oxygen and nitrogen species by repetitive negatively pulsed helium atmospheric pressure plasma jets propagating into humid air, S. A. Norberg, E. Johnsen, M. J. Kushner, Plasma Sources Science and Technology, VoL. 24, No. 3 (2015)

[53] Gas temperature effect on discharge-mode characteristics of atmospheric-pressure dielectric barrier discharge in a helium-oxygen mixture, W. S. Kang, H.-S. Kim, S. H. Hong, IEEE Transactions On Plasma Science, Vol. 38, No. 8, 1982-1990 (2010)

[54] Characteristics of atmospheric-pressure non-thermal N2 and N2/O2 gas mixture plasma jet, D. Xiao, C. Cheng, J. Shen, Y. Lan, H. Xie, X. Shu, Y. Meng, J. Li, P. K. Chu, Journal of Applied Physics, 115, 033303 (2014)

[55] Fast surface hydrophilization via atmospheric pressure plasma polymerization for biological and technical applications, Hana Dvořáková, Jan Cech, Monika Stupavská, Lubomír Prokeš, Jana Jurmanová, Vilma Buršíková , Jozef Ráhel', Pavel St'ahel, Polymers 2019, 11, 1613; doi:10.3390/polym11101613

[56] Quantitative phase tomography of Arabidopsis seeds reveals intercellular void network, P. Cloetens, R. Mache, M. Schlenker, S. Lerbs-Mache, PNAS September 26, 2006 103 (39) 14626-14630